\documentclass{aa}

\usepackage{graphicx}

\usepackage{txfonts}

\usepackage{natbib}

\begin{document}

\title{Buoyant Bubbles in a Cooling Intracluster Medium
       I. Hydrodynamic Bubbles}

\author{A. Gardini \thanks{gardini@astro.uiuc.edu}
        }
\institute{Department of Astronomy, University of Illinois at Urbana-Champaign, Urbana, IL 61801
\and
National Center for Supercomputing Applications, Urbana, IL 61801}

\date{Received / Accepted}

\abstract
{}
{Over the past several years, numerous examples of X-ray cavities 
coincident with radio sources have been observed 
in so-called ``cool core'' clusters of galaxies.
Motivated by these observations, we explore the evolution and 
the effect of cavities on a cooling intracluster medium (ICM) numerically,
adding relevant physics step by step.}
{In this paper we present a first set of hydrodynamical, 
high resolution ($1024^3$ effective grid elements), 
three-dimensional simulations, 
together with two-dimensional test cases. 
The simulations follow the evolution of radio cavities, 
modeled as bubbles filled by relativistic plasma,
in the cluster atmosphere while the ICM is subject to cooling.} 
{We find that the bubble rise retards the development of a cooling flow
by inducing motions in the ICM which repeatedly displace the 
material in the core. Even bubbles initially set significantly far from 
the cluster center affect the cooling flow, although much later 
than the beginning of the simulation. The effect is, however, modest: 
the cooling time is increased by at most only 25\%.
As expected, the overall evolution of pure hydrodynamic bubbles is at odds
with observations, showing that some additional physics has to be 
considered in order to match the data.}
{}


\titlerunning{Buoyant Bubbles in a Cooling ICM}
\authorrunning{Gardini}

\maketitle

\section{Introduction}

Approaching the centers of rich clusters of galaxies, 
one often finds a remarkable increase in the X-ray luminosity 
together with a marked decrease in the temperature. 
Under the conditions present in these clusters, 
the radiative cooling time of the intracluster medium (ICM) 
becomes shorter than the cluster lifetime and a ``cooling flow'' is expected 
to occur, i.e. a subsonic flow of cooling gas toward the cluster center
\citep[see][for a review]{Fabian94}.
The absence of observations of a reliable repository for the cold gas
left some doubts regarding this picture, 
which were confirmed when a major revision was forced
by spectral measurements of the cores of the clusters.
High-resolution XMM/Newton \citep{Peterson01,Peterson03,Tamura01,Kaastra04} 
spectra and lower resolution ASCA \citep{Makishima01} 
and Chandra \citep{David01} spectra show 
that only a small fraction of the gas is cooling below 1-2 keV, 
whereas its temperature was expected to be as low as $10^6$~K (0.1 keV).
Hence it is now said that these clusters host a ``cool core'' instead
of a cooling flow, and therefore in the rest of the paper we will 
refer to them as cool core cluster.

Maintaining gas at keV temperatures for a period several times longer than
its cooling time requires one or more heating mechanisms.
Candidate heating mechanisms include thermal conduction by electrons in the
ICM \citep{Narayan01,Zakamska03,Voigt04}, reconnection of 
magnetic fields \citep{Soker90}, turbulent mixing \citep{Kim03},
and the injection of hot plasma by the active galactic nucleus
(AGN) hosted by the central galaxy.
In this paper we will focus on this last possibility, but we do not 
exclude the possibility that more than one mechanism is at work.

Recent observations suggest that almost all cool core clusters
host radio sources in their centers \citep{Eilek04}.
Often these sources are rather faint, and it is not clear
if a correlation exists between their radio power and the strength
of the expected cooling flow \citep{Voigt04,Kaastra04,Birzan04}.
However, about 71\% of the cD galaxies in cool core clusters
are radio-loud compared to only 23\% of non-cool core cluster
cDs \citep{Burns90,Ball93}, and this suggests a connection
between AGN activity and the presence of a cool core,
at least in clusters that host a cD.

The shapes of radio sources are determined by the interaction of the
jets with their surroundings. The central radio sources in cool core 
clusters mostly show a disturbed morphology: even when they host 
a radio-loud core, collimated jets exist only on kpc scales or below.
After this the plasma flow continues in a less collimated manner into the ICM,
and the lobes take the shape of large plumes or wide tails
\citep[see e.g.][]{Eilek04,Jetha05}.
A remarkable exception is Cygnus A, 
a typical Fanaroff-Riley II (FRII) source, 
whose jets travel straight for about 100 kpc and terminate in hot spots
at the far ends of the lobes \citep[see e.g][]{Barthel96}.

The interaction of a single jet or a pair of jets 
with the ICM has been studied extensively both theoretically 
\citep{Scheuer74,Blandford74,Smith83,Begelman89,Heinz98,Alexander02}
and through numerical simulations 
\citep[see e.g.][and references therein] 
{Clarke97,Rizza00,Reynolds01,Reynolds02,Krause03,Krause03a,Zanni03}.
These simulations, however, mostly apply only to FRII sources and to the
phase of AGN activity.
This work suggests that jets inflate a cocoon with radio plasma 
which encompasses both the jets and the AGN, and as the radio source evolves, 
this cocoon elongates in the jet's direction and assumes a cigar-like shape.
The long-term evolution of the radio sources has received less attention;
on the one hand, it has been argued that they should continue to expand 
subsonically while the cocoon remains overpressured with respect to the 
surrounding medium.
On the other hand, it was pointed out that the cocoon evolution
would instead be driven by buoyancy when its expansion speed 
became comparable to the rising buoyancy speed \citep{Churazov00}. 
A more detailed picture has been revealed by some recent numerical studies
\citep{Reynolds02,Basson03,Omma04a,Omma04b,Zanni05}:
after the jets are switched off, they show that lateral
instabilities bisect the cocoon in its central region,
while its remaining parts start to rise in the cluster atmosphere,
due to residual momentum and to buoyancy.

Depressions of the X-ray flux coincident with AGN radio lobes
have been observed in cool core clusters
and also in some groups and galaxies by ROSAT and Chandra telescopes.
They are usually interpreted as cavities in the ICM, produced and
filled by the radio-emitting plasma injected by the central AGN
as described above.
A systematic study of these systems is provided by \citet{Birzan04}.
Usually, cavities occur in pairs, and the cavities in the same pair
are in opposite directions and at about the same distance 
with respect to the cluster center. 
In general, they also have similar dimensions,
measuring up to some tens of kpc across, 
and are located some tens of kpc away from the cluster center. 
However, a pair of huge cavities, roughly 200 kpc in diameter, 
has been detected by \citet{McNamara05} in MS0735.6+7421.

Cavities are sometimes surrounded by bright rims of denser and cooler gas
\citep[see e.g. A2052:][]{Blanton01}, 
which presumably has been displaced, uplifted or entrained by the
hot plasma during the formation of the cavity or its subsequent motion.
In some cases weak shocks are observed surrounding the cavities 
in clusters
(Perseus: Fabian et al., 2006, 2003a;
Hydra A: Nulsen et al., 2005b;
MS0735.6+7421: McNamara et al., 2005)
and in M87 \citep{Forman05},
but so far strong shocks have been detected only in Centaurus A 
\citep{Kraft03}. A weak shock has also been reported around the radio source
in Hercules A \citep{Nulsen05a}, which apparently does not host cavities,
and spectral evidence of a shock is reported in A478 \citep{Sanderson05}.
A bow shock in front of radio lobes may have been detected in Cygnus A
\citep{Smith02,Balucinska05}.
In the case of a weak shock, it can be assumed that
the radio-emitting plasma in the cavities 
is almost in pressure equilibrium with the surrounding ICM.
Thus, due to their low density, the radio lobes will rise by buoyancy 
in the cluster atmosphere \citep{Gull73}, 
similarly to the remnants of the cocoon in the models described above. 

Chandra X-ray images also show cavities that are not coincident with
bright radio lobes, for example in Abell 2597 \citep{McNamara01},
Perseus \citep{Fabian00}, and Abell 4059 \citep{Heinz02}.
These structures are referred to as ``ghost cavities''.
The relativistic electrons in the radio lobes are expected
to lose enough energy via synchrotron emission to become invisible
in the radio after 50 to 100~Myr.
If this interpretation is correct,
the ghost cavities are buoyantly rising relics of a radio
outburst that ended at least 50 to 100~Myr ago \citep{Soker02}.

The commonly observed ``cold fronts'' discovered by Chandra can also provide
hints regarding the effect of AGN on the intracluster medium.
Cold fronts are sharp discontinuities in
X-ray surface brightness marking boundaries between hotter and
colder masses of gas in the ICM. The density and temperature jumps
inferred for cold fronts through deprojection analysis are consistent
with continuous pressure changes across the fronts, showing that
the discontinuities are not shocks \citep{Markevitch00}.
While cold fronts are usually explained as remnants of past merger
events, they have been observed also in apparently regular clusters
such as RXJ~1720.1+2638 \citep{Mazzotta01}, Abell 1795 \citep{Markevitch01},
2A~0335+096 \citep{Mazzotta03}, and MS~1455.0+2232 \citep{Mazzotta01b}.
The gas sloshing induced by the motion of a pre-existing cavity
may explain these observations \citep{Mazzotta03}.

The H$\alpha$ filaments observed in the core of the Perseus
cluster also appear to be related to the upward motion of the cavities.
In particular, some filaments seem drawn up in a laminar flow in the wake 
of a ghost cavity, also tracing well-defined arcs that resemble
a circulation flow.
This suggests that the medium is not turbulent and has 
a non-negligible viscosity \citep{Fabian03b}.

Studies of the heating of the ICM through energy injection by AGN
have also been performed using analytical and semi-analytical models
\citep{Begelman01,Ruszkowski02, Bohringer02, Churazov02, Kaiser03,
Bruggen03a, Mathews03, Mathews04, Mathews06, Roychowdhury04},
which, however, describe these systems in very general terms.
This is due to the fact that the dynamics of the hot plasma in the ICM 
is a very complex problem, owing to its intrinsic three-dimensionality,
and to the chaotic interactions between plasma and ICM.
While analytical models of buoyant bubbles have been produced
\citep{Gull73,Churazov00,Soker02,Kaiser05},
the main tool of investigation consists of numerical simulations.
The interaction of the radio sources with the ICM 
has been simulated so far in different ways.
Their formation is often modeled 
by injecting energy or hot plasma directly into the ICM
\citep{Quilis01,Brighenti02a,Brighenti02b,Brighenti03,
Bruggen02a,Bruggen02b,Bruggen05, Bruggen03b,
Ruszkowski04a, Ruszkowski04b, DallaVecchia04, Jones05, Vernaleo06},
rather than reproducing detailed AGN jets as discussed above.
This approach enables one to study repeated bursts of activity,
whereas jet simulations at high resolution are usually unable to cover 
long evolutionary times.
In a different approach, the phase of buoyancy is reproduced
by modelling the radio sources as bubbles of hot plasma 
\citep{Churazov01,Bruggen01,Saxton01,Robinson04,Reynolds05},
thus neglecting the phase of formation of cavities, and starting
from a configuration in pressure equilibrium similar to observations. 
A review of simulations of hot plasma in clusters is given in \citet{Gardini04}.

We aim to investigate the properties of the ICM, the dynamical evolution 
of the radio cavities and their impact on the environment, by means 
of simulations of increasing complexity and realism.
In this paper we show the results of a first set of three-dimensional (3D)
hydrodynamical simulations in which cavities are modeled 
as buoyant bubbles in the atmosphere of a cool core cluster,
together with two-dimensional (2D) test cases.
With respect to previous similar numerical experiments, we achieve a rather 
high effective numerical resolution (1024$^3$ grid elements), and account
at the same time for the cooling of the gas.
Similar simulations have been performed in three dimensions 
only by \citet{Reynolds05}, who focus on the effect of ICM viscosity 
on the bubble evolution and therefore neglect cooling.
This work deals both with the dynamics of the bubbles and their effect 
on the cooling of the ICM. We also stress the reliability 
and robustness of the results, in order to use them as reference for future 
simulations. 

The paper is organized as follows: in \S 2 we describe 
the characteristics of the simulations and of the code. 
In \S 3 we describe the evolution of the simulations,
while the effects of bubbles on the cool core are discussed in \S 4. 
Finally, in \S 5 we summarize our results and discuss future developments 
of this work.

\section{Characteristics of simulations}

We defined the cluster parameters in order to resemble 
the observed characteristics of Abell 2390, 
a massive galaxy cluster with a significant cool core. 
We model the gravitational potential according to the \citet{Navarro97}
density distribution for dark-matter halos, 
\begin{equation}
\rho(r) = \rho_s \left[ 
    \left( r \over r_s \right)  
    \left( 1 + {r \over r_s} \right)^2 \right]^{-1}  
\end{equation}
and keep it fixed during each run.
We set the scale radius $r_s$ to 520 kpc, while the scale density
$\rho_s = 7.07 \times 10^{-26}$ g/cm$^3$ is derived by imposing 
the requirement that the virial mass ($M_{200}$)
of the cluster be equal to $1.7 \times 10^{15}$ M$_{\textrm{\sun}}$,
after assuming the Hubble parameter $h = 0.7 $ and redshift $z = 0$; 
the corresponding virial radius is then $r_{200}\sim 1.72 h^{-1}$ Mpc.

The initial gas temperature profile is shown in Figure 1; 
it is constructed as an analytic adaptation of the profile of Abell 2390 
proposed by \citet{Zakamska03}, which in turn reproduces by eye the temperature 
profile obtained by deprojection from Chandra observations by \citet{Allen01}. 
The expression for $T(r)$ is
$$
T(r) = \left ( T_{out} - (T_{out}-T_{in}) 
                   \exp (-r^2/(2\sigma(r)^2)) \right ) \times \nonumber 
$$
\begin{equation}
           \exp \left ( {1\over2}  \left ( \mu  (\log(r) - \log(r_s)) -
           \sqrt{  \mu^2(\log(r) - \log(r_s))^2 + 0.004 } \right ) \right )
\end{equation}
which requires that the temperature rise
from $T_{in}=4$ keV at the center up to $T_{out}=11$ keV at the scale radius.
Beyond the scale radius we assume that the gas temperature drops like
$T \propto r ^\mu, \mu=-0.4$, as suggested by the results of \citet{DeGrandi02}.
The definition of $\sigma$ is
\begin{equation}
\sigma(r) = a r + b
\end{equation}
with
\begin{equation}
a = \left ( {r_s\over 3} - {r_{ave}\over {\sqrt{-2\log0.5}}} \right ) 
       / (r_s - r_{ave})
\end{equation}
and 
\begin{equation}
b = r_s \left ( {1\over 3} - a \right )
\end{equation}
where the radius $r_{ave} = 80 $ kpc 
encompasses the part of the cluster where $T< (T_{out}+T_{in})/2$. 

\begin{figure}
\centering
\resizebox{\hsize}{!}{\includegraphics{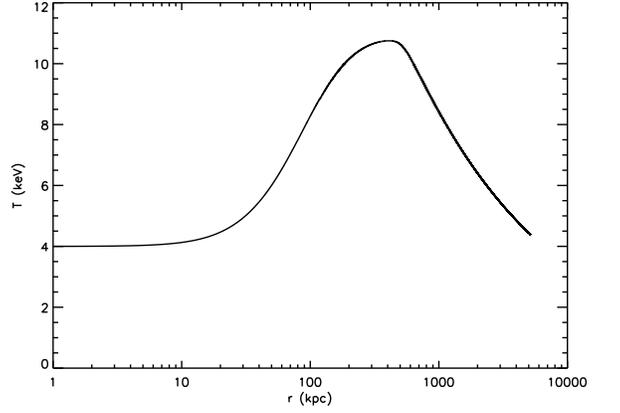}}
\caption{
The temperature profile of the simulated cluster. 
The temperature rises from about 4 keV at the center up to 11 keV 
at the scale radius, before decreasing in the outskirts.
The profile is derived from the data of \citet{Allen01} 
as in \citet{Zakamska03} for Abell 2390.
Beyond the scale radius the profile goes as $T\propto r^{-0.4}$, 
as suggested by the results of \citet{DeGrandi02}.}
\label{fig1}
\end{figure}

The gas density profile is obtained by solving the equation of hydrostatic
equilibrium with the condition that the gas fraction inside the
virial radius be $f_b = 0.15$, approaching the cosmic value 
\citep[see e.g.][]{Spergel03}.
In Figure 2 we show the profile of the cooling time 
$t_{\rm cool} = - (d\, \ln T / dt)^{-1}$
for the initial conditions according to the cooling function 
of \citet{Sutherland93} used in the code.
It is worth noting that the region of the cluster where 
$t_{\rm cool} \le 1$ Gyr spans only $r \le $ 12 kpc.
The adiabatic index of the ICM is set to $\gamma=5/3$, assuming
a monatomic perfect gas.
We tested the equilibrium of the initial configuration of the ICM
by running a 2D simulation without cooling or bubbles. 
After $t=500$ Myr, readjustments in the physical quantities 
(density, pressure, etc.) amounted at most to 1\%.

\begin{figure}
\centering
\resizebox{\hsize}{!}{\includegraphics{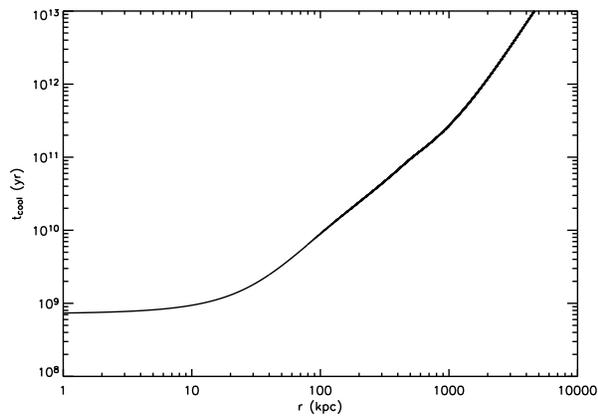}}
\caption{
The value of the cooling time 
$t_{\rm cool} = - (d\, \ln T / dt)^{-1}$
as a function of the radius.}
\label{fig2}
\end{figure}

The rising radio lobes are simulated as a pair of spherical bubbles in the ICM, 
placed symmetrically with respect to the cluster center. 
Inside the bubbles, we set the gas density and pressure respectively
to 1/100 of and the same value as the ICM's quantities at the same radius;  
we also impose $\gamma = 4/3$ in the bubble material 
because the plasma is relativistic.
We did not attempt to include the ICM displaced from these cavities
during the phase of bubble formation.
In the following we will refer to the two gas components which had
$\gamma=5/3$ and $\gamma=4/3$ respectively as ICM and bubble plasma.

The different characteristics of the 3D runs are summarized in Table 1.
The letters CO (cooling only) identify simulations performed without bubbles.
In simulations with bubbles, the initials SB (single bubble pair) are
followed by the radius $R_b=10$ kpc of the bubbles and the distance $d_b$ 
of their centers from the center of the cluster.
The increasing computational cost did not permit us to run the simulations for
their entire cooling time, but they cover a significant fraction of it, 
up to $t=250$ Myr.
A set of simulations with $R_b=20$ kpc halted at earlier stages of evolution
and has not been included in this analysis. 
All of these simulations are performed on a Cartesian grid.
We also created a number of 2D simulations in cylindrical
coordinates to test and extend the 3D results. 
Because of the symmetry of the system, it can be noted that the same amount 
of information extracted by a 3D simulation could be obtained through 
a faster and computationally cheaper 2D counterpart. 
We noticed, however, that the two bubbles in these 2D runs evolve in a 
similar but not fully symmetric manner. This is probably due to a greater
amplification of rounding errors in this particular system of coordinates
with respect to the Cartesian grid. The evolution of the bubbles and 
their effect on the medium are nonetheless in agreement with the 3D cases
until they are available, and this encouraged us to trust the 2D results,
at least as checks or hints of the overall evolution of the system.
We will not discuss the 2D runs extensively but we will report their 
results when necessary. 

\begin{table}
\begin{minipage}[t]{\columnwidth}
\begin{center}
\caption{Parameters of the 3D simulations.\label{tbl-1}}
\renewcommand{\footnoterule}{}
\begin{tabular}{crrrrrr}
\hline\hline
Name & 
$L_{box}$\footnote{The box size in kpc} &
$\Delta x$\footnote{The maximum resolution in pc} &
$R_b$\footnote{The radius of the bubbles in kpc} &
$d_b$\footnote{The distance of the center of the bubbles from the center 
of the cluster in kpc} &
$E_{int}$\footnote{The internal energy of the bubbles in units of $10^{57}$ 
erg} &
$t_{final}$\footnote{The time of the final output of the simulation in Myr} \\
\hline
CO & 4800 & 4688 & -  & - & - & 500 \\
CO2 & 600 & 586 & -  & - & - & 250 \\
SB1011 & 600 & 586 & 10 & 11 & 337 & 250\\
SB1020 & 600 & 586 & 10 & 20 & 281 & 250\\
SB1030 & 600 & 586 & 10 & 30 & 235 & 250\\
\hline
\end{tabular}
\end{center}
\end{minipage}
\end{table}

The 3D computational box is centered on the cluster center and in general
spans 600 kpc; its dimensions are a good compromise between 
resolution and the need to avoid spurious boundary effects.
Because we chose reflecting boundary conditions to conserve mass and energy, 
the box size has to be large enough that the cooling flow 
and the bubbles are unaffected.
We achieve a maximum resolution of $\sim 586$ pc, 
comparable to the electron mean free path in the cluster core, 
when magnetic fields are neglected \citep{Sarazin88}. 
The CO model, which uses a box size of 4800 kpc and a resolution of
$\sim 4.7$ kpc, is a test case for the development of the cooling flow
in a larger box. 

The cooling function of \citet{Sutherland93} is implemented in the code;
we shut off cooling below a temperature of 10$^4$ K 
and assume an ICM metallicity of $\sim  0.3 Z_\odot$.

All simulations are performed using FLASH \citep{Fryxell00}, 
a parallel Eulerian hydrodynamic code which implements 
the piecewise-parabolic method (PPM) of \citet{Colella84} on an adaptive mesh. 
Grid cells are grouped in blocks, which are refined or derefined 
according to the maximum value of the second derivative of pressure or density 
\citep{Loehner87}. In 3D simulations the volume is divided into $4^3$ blocks 
at the coarsest level of refinement, and each block contains $8^3$ cells. 
Up to 6 levels of refinements are allowed, with a factor of two refinement
between levels. 

FLASH manages mixtures of gases using the method of \citet{Colella85}
to handle general equations of state. For the purposes of this work
it is worth noting that each fluid element in the simulation is composed
of a fraction of both the ICM and the bubble plasma,
and that all of the physical quantities, such as density, temperature 
and pressure, are evaluated on the mixture of the two species. 
In particular to each cell it is assigned a weighted average adiabatic
index $\gamma$ as
\begin{equation}
{1\over {(\gamma - 1)}} = \sum_i {X_i\over {(\gamma_i - 1)}}
\end{equation}
where $X_i$ is the mass fraction of the $i$th species and
$\gamma_i$ its adiabatic index.
This left us with two possible problems: on one hand, 
mixing is not properly treated in these simulations
because the diffusion of species is not explicitly accounted for. 
On the other hand, the temperature in the plasma elements drops quickly 
as they are polluted by the ICM, but the relativistic value $\gamma=4/3$ 
of their adiabatic index does not change.
We address the possible effects of these inconsistencies in the following 
sections.

\section{Evolution of simulations}

\subsection{The development of the cooling flow}

The simulations named CO and CO2 in Table 1 allow the ICM to cool,
in the absence of bubbles or any source of heating. They reproduce the effect 
of cooling on the ICM starting from the equilibrium configuration
assumed in the initial conditions. The simulations are identical except
for the box size and the corresponding resolution.

Figure 3 shows the evolution of temperature $T$, electron density $n$,
entropy $s = T/n^{2/3}$, and pressure $P$ in the central part of the cluster
due to cooling.
Profiles are obtained for the CO2 (solid lines) and CO (dashed lines)
simulations by averaging on density and assuming a bin interval of 3 kpc.
The temporal evolution for the CO2 simulation is represented 
by the line thickness, which decreases with time: 
profiles are computed at intervals of 50 Myr
from the initial conditions up to $t = 250$ Myr.
The dashed lines represent the profiles of the same quantities in the CO
simulation to $t=250$ Myr and $t=500$ Myr; at $t=250$ Myr we also added 
squares to the dashed line to facilitate visual comparison. 
The figure shows that the profiles at $t=250$ Myr are practically coincident 
to the limit of resolution; this implies that the finite size of the box
does not induce spurious effects in the CO2 simulation. 

\begin{figure}
\centering
\resizebox{\hsize}{!}{\includegraphics{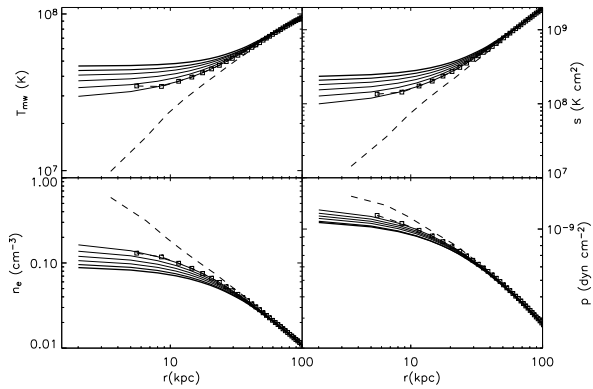}}
\caption{
The evolution of the profiles of some of the ICM quantities due to cooling,
in the absence of cavities. 
The plot shows the evolution of temperature $T$, electron density $n$,
entropy $s = T/n^{2/3}$, and pressure $P$ in the central part of the cluster.
Profiles are obtained from the CO2 (solid lines) and CO (dashed lines)
simulations by averaging over radial bins, assuming a bin interval of 3 kpc.
The time evolution for the CO2 simulation is represented by the line thickness,
which decreases with time. 
Profiles are computed with an interval of 50 Myr from the initial conditions 
up to $t = 250$ Myr.
The dashed lines represent the profiles of the same quantities in the CO
simulation at $t=250$ Myr and $t=500$ Myr. Squares have been added to 
the profiles of the CO simulation at $t=250$ Myr in order to aid
comparison with the CO2 case.}
\label{fig3}
\end{figure}

As expected, these runs show the establishment of an homogeneous cooling flow 
which evolves in a runaway manner \citep{White87}. 
Because of the long cooling time, it is not surprising that the temperature 
of the core drops only to $T = 3 \times 10^7$ K after the first 250 Myr. 
Using a 2D simulation with the same resolution 
as the CO2 case, we checked that the central temperature
falls to the minimum value between $t=390$ and 400 Myr.   
We will use the profiles obtained in the CO2 case as references
for the same quantities computed in the presence of bubbles. 

\subsection{The rise of the bubbles}

The rise of spherical bubbles of plasma in a cluster atmosphere
has been described in many papers \citep[e.g.][]{Churazov00}.
We observe that the evolution of bubbles in the ICM 
resembles quite closely the dynamics of powerful explosions 
in the earth's atmosphere, or more generally of ``thermals'', 
which are spherical bubbles of hot air in pressure equilibrium 
with the surrounding colder air. 
In the latter case, the main difference resides in the fact 
that thermals are only slightly hotter and less dense than the surrounding air; 
this produces a rather different behavior, and thermals are able
to rise in the atmosphere much higher than their initial
diameter, expanding linearly with time \citep{Turner73}.
Studies of thermals can hence adopt the Boussinesq approximation 
in computations and perform experiments using liquids.
Nevertheless, some of the behavior of thermals, such as their internal motions,
presents remarkable similarities with the behavior of plasma bubbles.
The behavior of spherical bubbles in water also resembles,
for a proper choice of parameters, the behavior of plasma bubbles 
\citep[e.g.][]{Walters63}. 

The evolution of the 3D simulations with bubbles is shown in Figures 4 to 9
through maps constructed by cutting the simulations along one of the planes
of symmetry which contain the centers of the cluster and the bubbles.
We concentrate on the simulation SB1011 as representative of the sample,
and will refer to the other 3D simulations only when their behavior 
is quite different with respect to it.

We paid some attention to the first stages of the evolution, namely 
the starting of the bubble from rest and the generation of sound waves.
This phase is usually neglected in the literature as transient and unrealistic
\citep[see however][]{Churazov02} despite the fact that 
sound waves are often clearly visible in figures. 
In Figure 4 we show the evolution of the pressure and velocity fields during 
the first 10 Myr. 
The plots show how the bubble plasma oscillates vertically and generates 
sound waves, while motions are driven into the adjacent ICM at the same time.
At the time of the last snapshot, a vortex circulation directed upward along 
the bubble axis and downward on the bubble surface \citep{Hill94,Batchelor67}
is established in the plasma and in the surrounding ICM.
This kind of circulation enables the bubble to rise steadily \citep{Turner73}.

\begin{figure*}
\centering
\includegraphics[width=17cm]{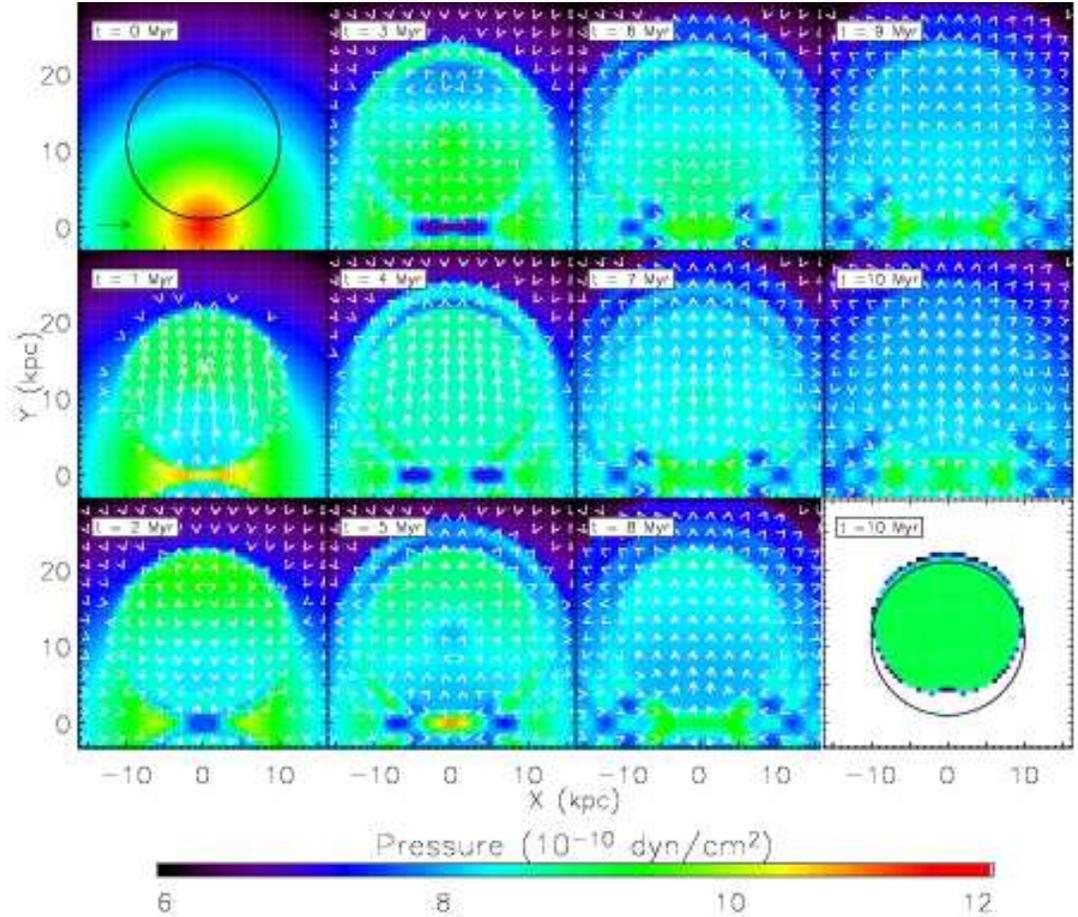}
\caption{
The evolution of the pressure field from $t=0$ up to $t=10$ Myr
for the simulation SB1011 in the region around the bubble, 
with the velocity field superimposed.
The first picture shows the border of the bubble as well as 
a velocity vector corresponding to $v = 1000$ km/s.
The last snapshot shows the region occupied 
by the fluid elements containing at least 90\% hot plasma 
at $t=10$ Myr, with the initial position of the bubble superimposed on it.
The bubble rises $\sim 2$ kpc in the first 10 Myr.}
\label{fig4}
\end{figure*}

During the rise that follows the initial transient phenomena, 
each bubble is affected by instabilities on its surface 
due to lack of surface tension. In particular it will suffer
Rayleigh-Taylor (RT) instabilities on the top and Kelvin-Helmoltz (KH)
instabilities on its flanks.
In the end, however, it is the vortex circulation itself which mostly
affects the bubble. Indeed, in an ideal case a stagnation layer 
should be established in the ICM below the bubble, 
but in reality, as the bubble rises, the fall of pressure
at its bottom draws the ICM upward, where it is entrained in the 
vortex circulation, thus acting as an effective RT instability.

\begin{figure*}
\centering
\includegraphics[width=17cm]{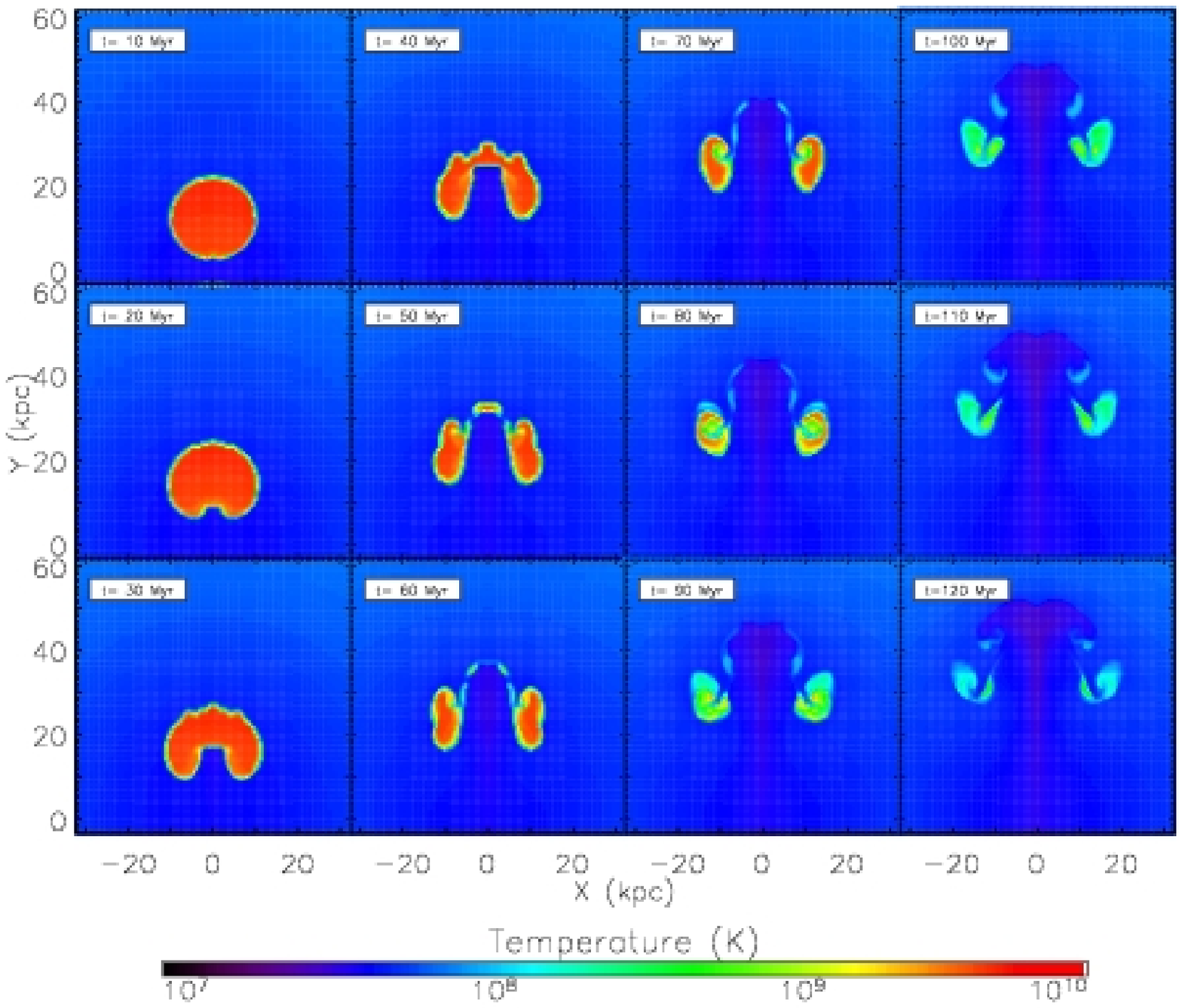}
\caption{
Map of the temperature in the SB1011 simulation showing
the evolution of the plasma bubble in the first 120 Myr.
The pictures show clearly the effect of the entrained cold gas,
which penetrates and finally bisects the ICM bubble, 
leaving a vortex ring.}
\label{fig5}
\end{figure*}
 
The evolution of the rising bubbles is depicted in Figure 5 
with intervals of 10 Myr up to $t=120$ Myr and in Figures 6 to 8 with
intervals of 50 Myr up to $t=250$ Myr. 
As long as the bubble rises, a trunk of denser material is entrained 
in its wake, which penetrates it from below and gives it a mushroom shape. 
As expected in the development of RT instabilies, the uplifted material 
rises faster and eventually bisects the bubble, which then assumes
the shape of a rotating torus (or vortex ring). 
It is worth noting that the ICM entrained by the bubble cools 
by expansion because of the lower pressure, 
and at some times it is the coldest material in the cluster. 
As the evolution goes on, the lower part of the trunk 
falls back to the cluster center, while the upper part remains
at large radii as long as it is driven in the vortex circulation.

\begin{figure}
\centering
\resizebox{\hsize}{!}{\includegraphics{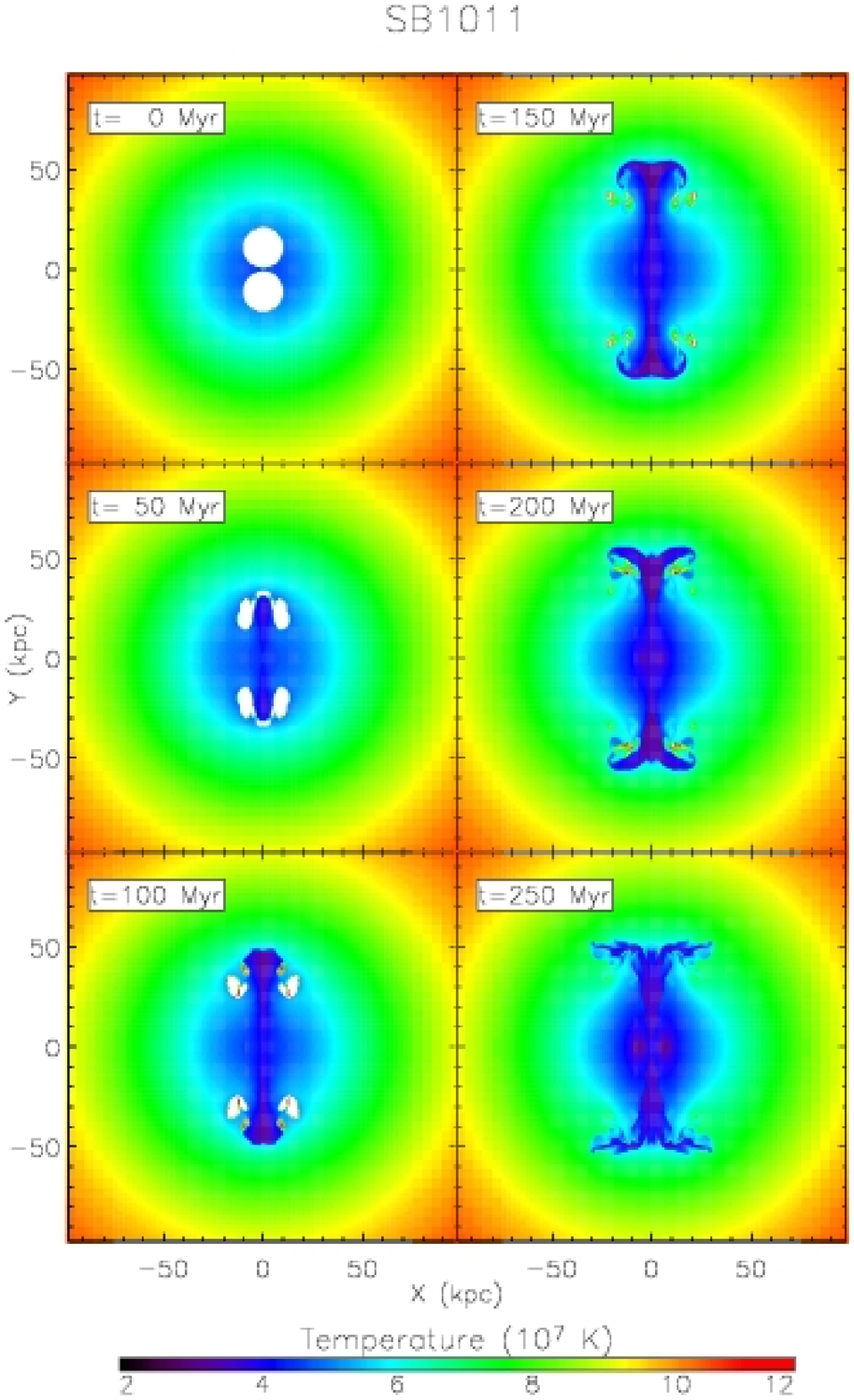}}
\caption{
The evolution of the ICM temperature in the SB1011 simulation 
from the beginning of the simulation up to $t=250$ Myr.
The bubble plasma seems almost completely diffused into the ICM in last
snapshot, but it is still preserved in remnants of the vortex
ring that do not intersect the plane of the image.
The pictures show the entrainment of cold gas along the vertical axis,
the establishment of a cold cap at large radii, the fall back of
the cold gas and the development of the cooling flow.}
\label{fig6}
\end{figure}

\begin{figure}
\centering
\resizebox{\hsize}{!}{\includegraphics{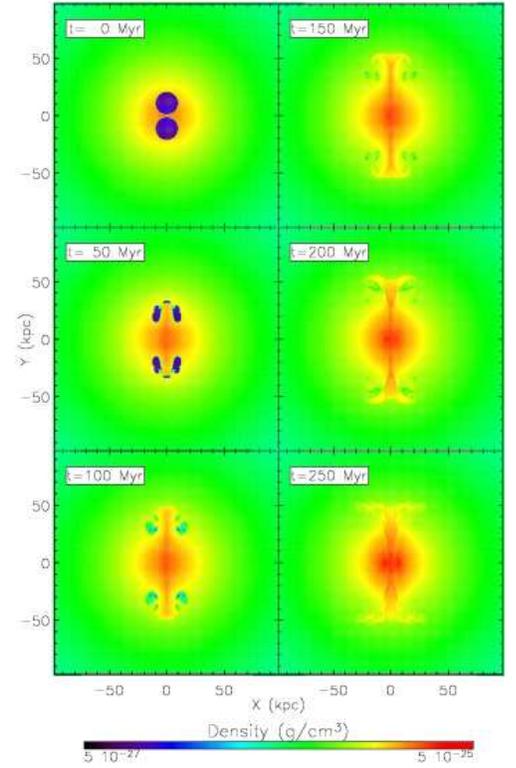}}
\caption{
The evolution of the density in the SB1011 simulation up to $t=250$ Myr.
The snapshots show the denser material that is uplifted by the vortex 
circulation and later sustained at large radii by the buoyancy of the
vortex ring.}
\label{fig7}
\end{figure}

\begin{figure}
\centering
\resizebox{\hsize}{!}{\includegraphics{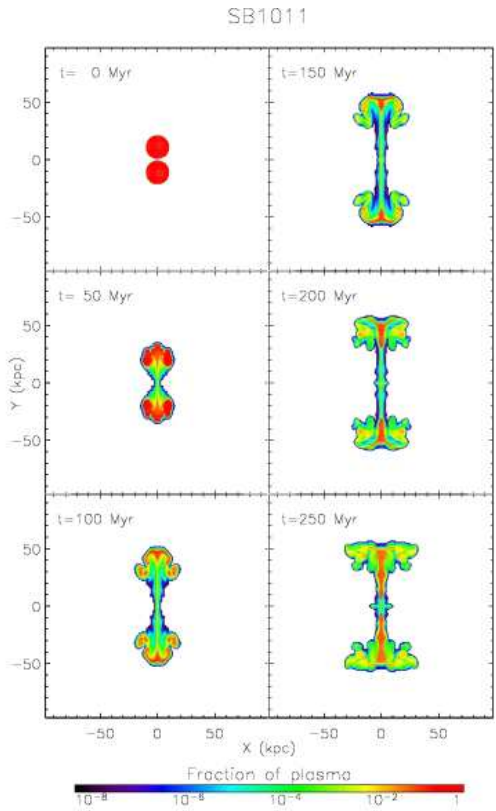}}
\caption{
The fraction of bubble plasma in the fluid elements in the SB1011 simulation
up to $t=250$ Myr. The comparison with Figure 6 and 7 shows that as the
simulation goes on, most of the bubble plasma is indeed mixed with the ICM
to low temperature and high density, and under these conditions it should 
no longer be treated as relativistic.
The reader should be careful to the logarithmic colour scale.}
\label{fig8}
\end{figure}

\begin{figure}
\centering
\resizebox{\hsize}{!}{\includegraphics{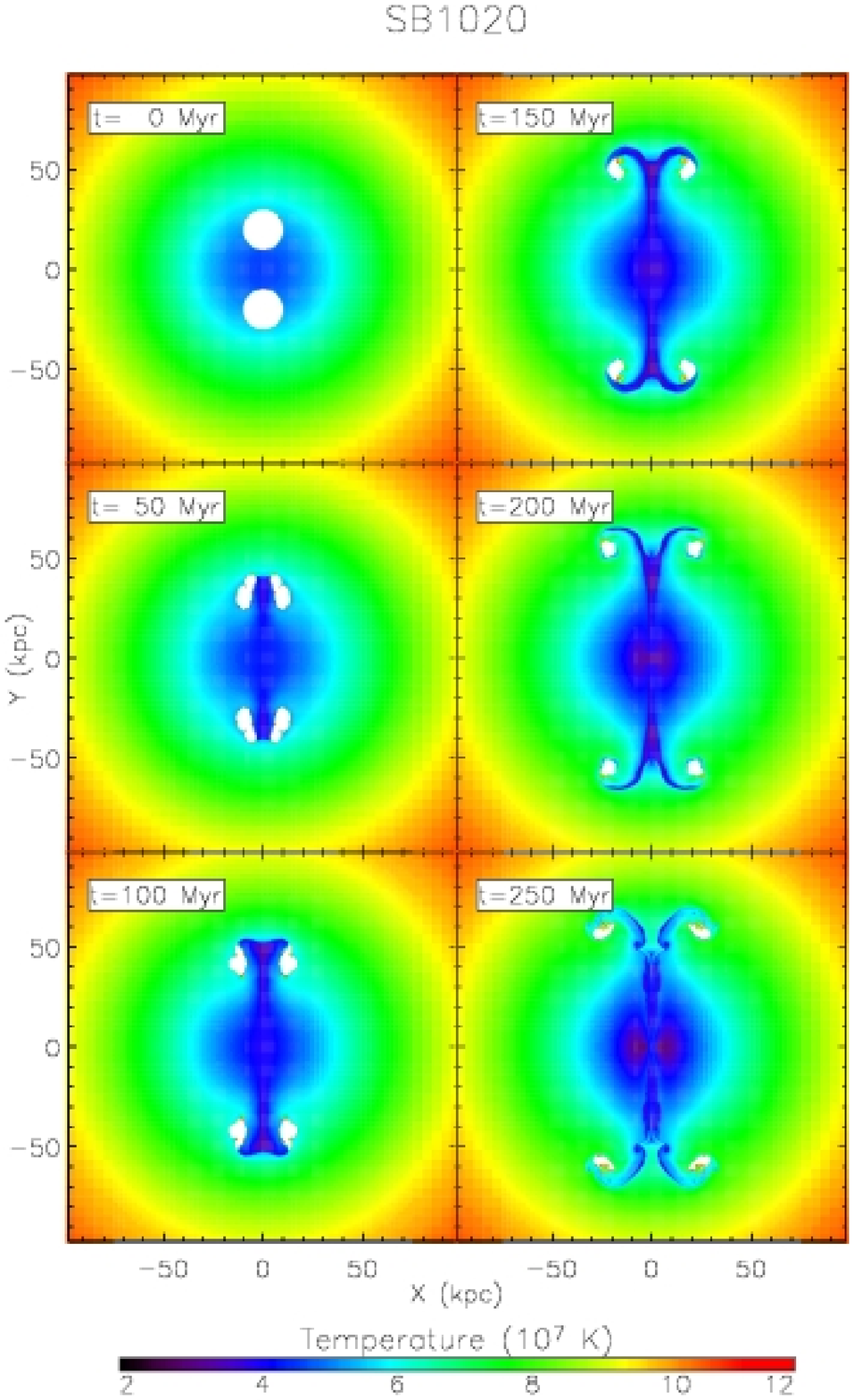}}
\caption{
The evolution of the temperature in the SB1020 simulation.}
\label{fig9}
\end{figure}

Defining the borders of the bubbles becomes a harder task
as the simulation proceeds.
Figure 8 shows the fraction of the bubble plasma in the fluid elements.
Inspection of these values shows clearly 
that all of the hot plasma has mixed completely with the ICM after 100 Myr. 
In particular, the plasma picked up along the bubble axis by the rising ICM
cools rapidly to low temperature due to the high density of the entrained ICM
and should no longer be treated as relativistic.
In the vortex ring, the plasma is also mixed with the ICM, but because 
of the lower density of the polluting ICM, 
the mixture still preserves buoyancy and a high temperature. 

As stated above, 
despite its relevance for the evolution of the bubbles and the ICM dynamics,
mixing is not properly treated in these simulations
because the diffusion of species is not explicitly accounted for. 
We address this point by checking that the amount of mixing remains stable
with respect to increases in resolution. We ran a 2D simulation with the same
characteristics as the SB1020 case and compared it with an identical
2D simulation with two additional levels of refinement.
We found that both simulations show the same amount of mixing 
up to $t=250$ Myr, despite the fact that the plasma distribution 
shows narrower features in the second case, as expected.
We consider that the mixing should eventually increase in the 2D simulations
due to their enhanced sensitivity to instabilities; hence the fact
that the amount of mixing remains the same is a proof of its robustness.

The vortex rings evolve similarly in the different simulations.
In all of the models the rings are affected by instabilities
at regular distances along their length that act to break
the rings into smaller sections. 
It is worth noting that the positions of the break points along the rings
are symmetric with respect to the principal vertical planes of the system,
and hence the instabilities do not reflect the axial symmetry of the cluster
but the geometry of the computational grid.
In the SB1011 simulation, the rings are already broken in the last outputs
exactly along the plane of the maps, so that it seems in Figure 6 that only
some remnants are slightly hotter than the surrounding medium 
after $t=200$ Myr. However, this is not the case: the hot material
is concentrated outside of the plane of the figure.
In the SB1020 and SB1030 simulations, 
the vortex rings are not yet broken by the end of the run, 
but instabilities are effectively breaking them outside the plane of the image.
In all of the simulations the rings preserve high temperatures despite the fact 
that the fraction of bubble plasma inside them is less than 10\%.

We quantify the evolution of the bubbles in Figure 10, which shows 
the height of the centroid of the plasma material as a function of time.
The dashed lines describe the value of the height of the centroid 
of all of the plasma, while the solid lines depict the same
quantity for the plasma that is hotter than $T=8\times 10^7 $ K
(the maximum temperature of the ICM inside $r=80$ kpc).  
The thickness of the lines distinguishes among the different simulations.
The three simulations show remarkably similar behavior: 
the hotter material, which effectively defines the bubbles,
rises by $\sim 20 $ kpc in the first 100 Myr, 
and then decreases its speed until the end of the simulation. 
The average speed is then $\sim 200 $ km/s for the first 100 Myr, 
which can be compared to the speed of sound in the ICM, ranging 
from $\sim 1000$ km/s at the center to $\sim 1500$ km/s at $r=100$ kpc.
The differences between the values for the overall plasma and the hottest 
plasma at the beginning of the simulation are explained by the fact that, 
due to the finite grid size, a portion of bubble material is mixed with the ICM
already in the initial conditions. Initially this material is less
affected by buoyancy, and it is driven upward later in the bubble wake.
The rising of the trunk of cold gas, the sweeping of the plasma on the
bubble axis and the establishment of a cold cap above the ring,
are also well represented by the rise of the overall centroid higher than
the hot plasma. The succeeding decrease
depicts the falling back of the cold material along the trunk.

\begin{figure}
\centering
\resizebox{\hsize}{!}{\includegraphics{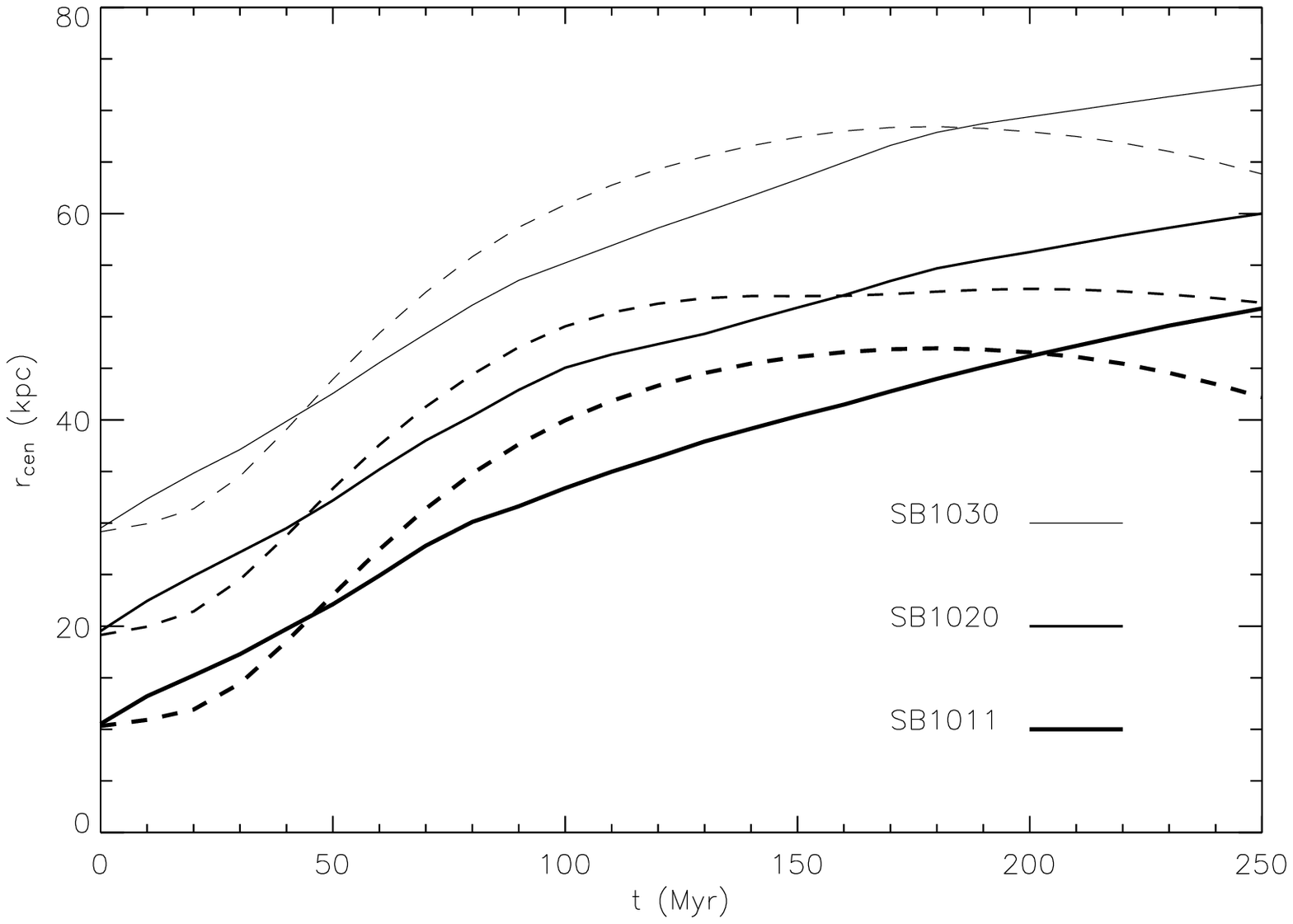}}
\caption{
The height of the centroid of the plasma material during each run
for the complete set of bubble simulations. 
The dashed lines describe the height of the centroid
of all the bubble plasma, while the solid lines depict the same
quantity for the plasma that is hotter than $T=8\times 10^7 $ K,
which is the maximum temperature inside $r=80$ kpc.}
\label{fig10}
\end{figure}

The centroid of the hot material still seems to rise steadily
in the last outputs of our simulations, but inspection of the 
fraction of plasma that is still hot, depicted in Figure 11, shows
that the plasma in the SB1011 simulation is rapidly cooling.
In the SB1020 and SB1030 simulations, it instead appears that 
the fraction of hot plasma is stable around 35\% in the final output. 

\begin{figure}
\centering
\resizebox{\hsize}{!}{\includegraphics{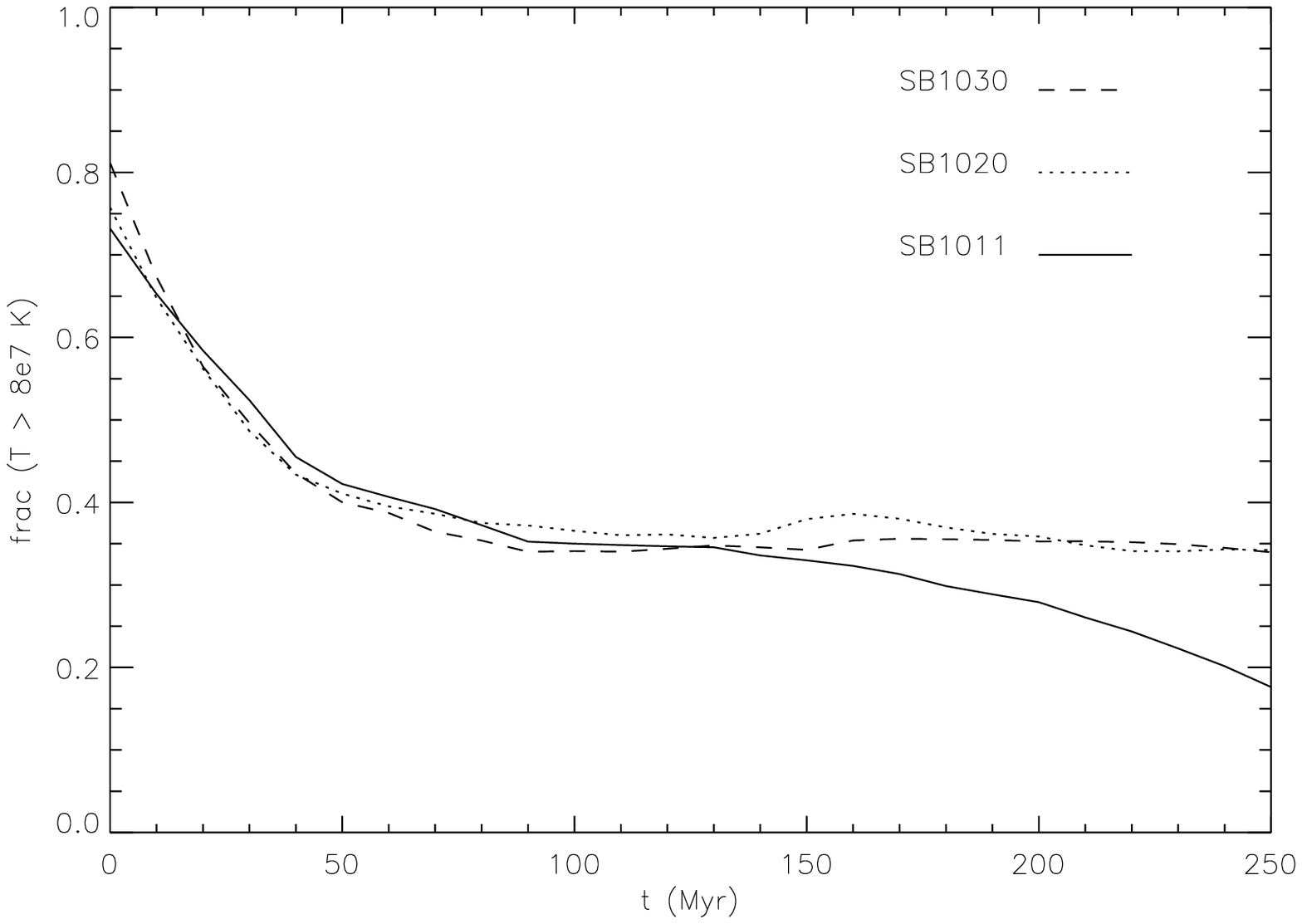}}
\caption{
The fraction of bubble plasma which maintains a temperature 
$T > 8\times 10^{7}$ K as a function of time. 
Even if the temperature of this plasma is lowered to the non-relativistic
regime, it is still hot and underdense enough to be driven by buoyancy.}
\label{fig11}
\end{figure}

To explore the later stages of the evolution, we turn again to 2D simulations
with the same parameters and resolution as the 3D cases. 
In general, these simulations show that material continues to fall along 
the trunks down to the cluster center, while the ICM already present there 
is displaced orthogonally and settles into a torus. 
The trend then reverses and the material in the torus falls back onto the core,
driving back upward the material that has fallen along the trunks.
Overall, this resembles a sort of oscillatory motion, 
but it becomes more turbulent as the simulation proceeds. 
We halt the simulations when some plasma element cools close to the minimum 
temperature and the computational time step becomes too short;
we can suppose that a runaway cooling similar to the homogeneus cooling flow
is established at that time. 
Vortex rings are preserved until the end of the run, deforming and slowly
rising for few kpc more. 

\section{Effects of bubbles on the cooling}

It has been observed that AGNs might heat the ICM and possibly
quench cooling flows in any of several different ways.
In particular, it has been predicted \citep[e.g.][]{Heinz98} 
and shown by simulations \citep{Zanni05,Bruggen05} 
that most of the jets' energy is transferred to the ICM 
through the bow shock and the sound waves pushed by the expanding cocoon.
These processes deposit energy into the ICM as the jets cross it, affecting 
the ICM at a distance from the radio source itself and possibly outside
the cool core, unless some viscosity or thermal conductivity is present 
\citep{Ruszkowski04a,Ruszkowski04b,Bruggen05,Fabian05}.
On the other hand, energy is still stored in the cavities as thermal energy
and in the ICM in potential form during the later phase of
evolution by buoyancy. Both these forms of energy are released during the
rise of the cavities: the first through $p dV$ work performed by the
cavities, which expand in order to reach or maintain pressure equilibrium
with the environment \citep[e.g.][]{Begelman01},
and the second when the surrounding medium falls in around the cavity 
to fill the space the bubble previously occupied 
\citep[e.g.][]{Churazov02,Birzan04}.
Work and thermal energy exchange also occur when portions of the ICM 
at different temperatures and pressures come into contact and mix, 
after being displaced by motions induced by the rise of the cavities 
\citep{Quilis01,Churazov02,DallaVecchia04}.
This mechanism becomes more effective in heating the cool core 
as more material is radially displaced.

Despite the presence of bubbles, our simulations show that
a cooling flow is established in the central regions of the cluster 
(Figures 6 and 9).
This is even more true as the initial location of the bubbles
is moved away from the center.
This is hardly surprising, given that bubbles are the outcome of
a single burst of energy injection while cooling is continously operating
in the ICM. Nonetheless, bubbles affect the ICM and can slow down its cooling 
effectively; in the following we investigate how and by what amount. 

Because of the dependence of gas emissivity upon density
(approximately $\epsilon\propto\rho^2 T^{-0.6}$),
cooling is enhanced by deposition and compression of material.
Thus, we observe that bubbles affect the cooling 
firstly by their mere presence, because of the removal from their volumes 
of some fraction of the cold ICM already stored in the core. 
Even if this can be regarded as an artifact of our setup, 
we expect that a similar effect is produced on the ICM 
by displacement and shock heating during bubble inflation.
Later, some amount of thermal energy is directly deposited in the ICM
through mixing with the bubble material, while stirring of the ICM
due to the bubble rise is expected to reduce the deposition of cooling 
material to the core. 
To investigate the effects of the gas motions on the cooling,
we performed a new run of the SB1020 case using a tracer fluid to mark
the fluid elements initially stored inside $r=10$ kpc;
we show in Figure 12 the evolution of their distribution. 
We observe that in the first stages of evolution,
the central material is stretched and uplifted in the bubbles' wake and
expands rather than being compressed. Later on, material indeed falls
onto the core, but because of the geometry, it accumulates directly at
the center rather than on the cold ICM, 
which is displaced instead into a torus. 
A similar analysis was performed also in the SB1011 case; its results
are similar to SB1020 up to $t=190$ Myr, when the run halts. 

\begin{figure}
\centering
\resizebox{\hsize}{!}{\includegraphics{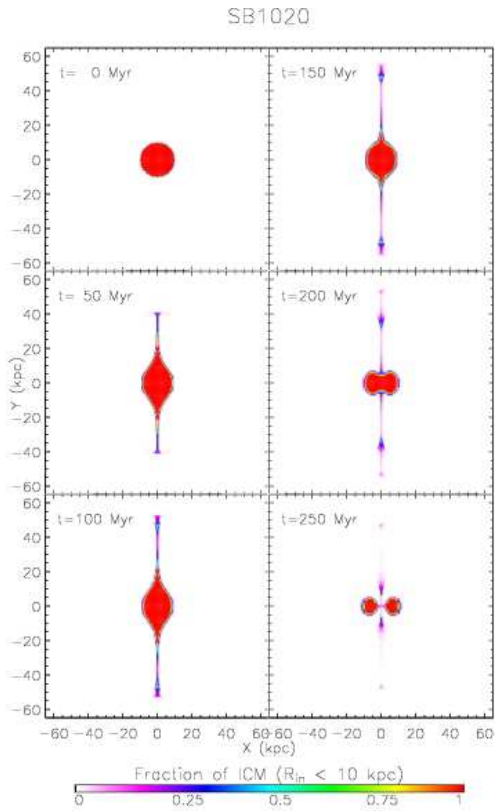}}
%\resizebox{\hsize}{!}{\includegraphics{f12.eps}}
\caption{
The evolution of the material initially inside $r=10$ kpc 
in run SB1020. It is first stretched by the rise of the bubbles and later
displaced into a torus around the center by the material falling along the 
trunks.}
\label{fig12}
\end{figure}

We measure the effects of the bubbles on the cooling by comparing
the temperature profiles with the profiles of the CO2 model.
This is done in Figure 13, which reports the difference $T-T_{CO2}$
from the initial conditions up to $t=250$ Myr with intervals of 50 Myr.
It must be remembered that, as depicted in Figure 3, in the absence of bubbles
the gas cools about $\Delta T\sim 2\times 10^7$ K during the first 250 Myr
of evolution, while the cooling is much faster at later times. 
Hence, a slight gain in the core temperature can be indicative 
of a significantly longer evolution of the cooling process.
The bubbles in simulation SB1011 have the greatest effect on the cooling, 
as expected, due to their proximity to the core of the cluster.
In this case the temperature is increased with respect to the CO2 simulation
in all of the region inside $r \sim 50$ kpc for most of the run. 
At $t=250$ Myr, the temperature difference reaches $T-T_{CO2} \sim 10^7$ K.
The most interesting result comes from the SB1020 simulation.
In this case the cooling in the center proceeds apparently 
as in the CO2 case up to $t\sim 200$ Myr. 
After this time, however, the central temperature drops
more slowly, and in the final snapshot $T-T_{CO2} \sim 6\times 10^6$ K.
This is due to the replacement of the cold material in the core
by the dense gas falling to the center along the trunks.
In the SB1030 simulation the core of the cluster instead appears completely
unaffected by the presence of bubbles up to the end of the run at $t=250$ Myr. 
However, because dense gas is clearly falling along the trunks at that time, 
a later heating as in SB1020 cannot be excluded.
Because of the fraction of plasma entrained in the core (Fig. 8),
we wondered if the higher temperature of the infalling material
was due to mixing. Therefore
we computed the profiles of the fraction of internal energy due to the 
bubble plasma as shown in Figure 14. At $t=250$ Myr, we found that the 
internal energy directly deposited in the ICM amounts at most at 2\% 
everywhere except in the shells hosting the vortex rings. 

\begin{figure}
\centering
\resizebox{\hsize}{!}{\includegraphics{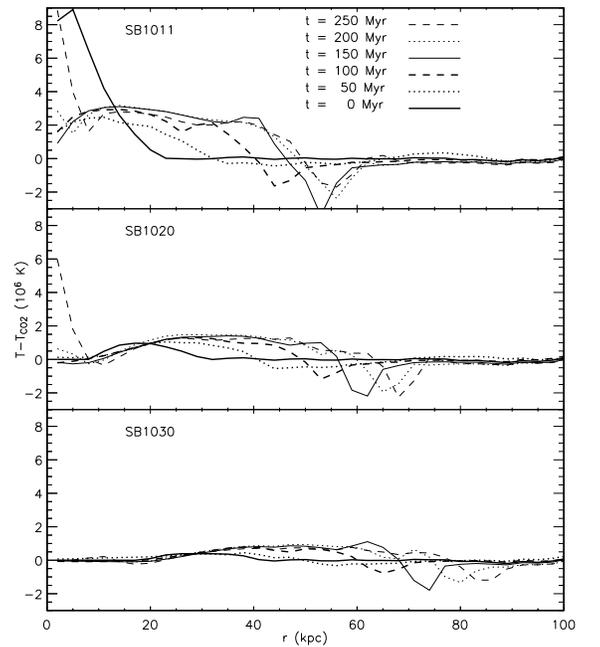}}
\caption{
The difference between the temperature profile of the simulations hosting
bubbles and the CO2 case.}
\label{fig13}
\end{figure}

\begin{figure}
\centering
\resizebox{\hsize}{!}{\includegraphics{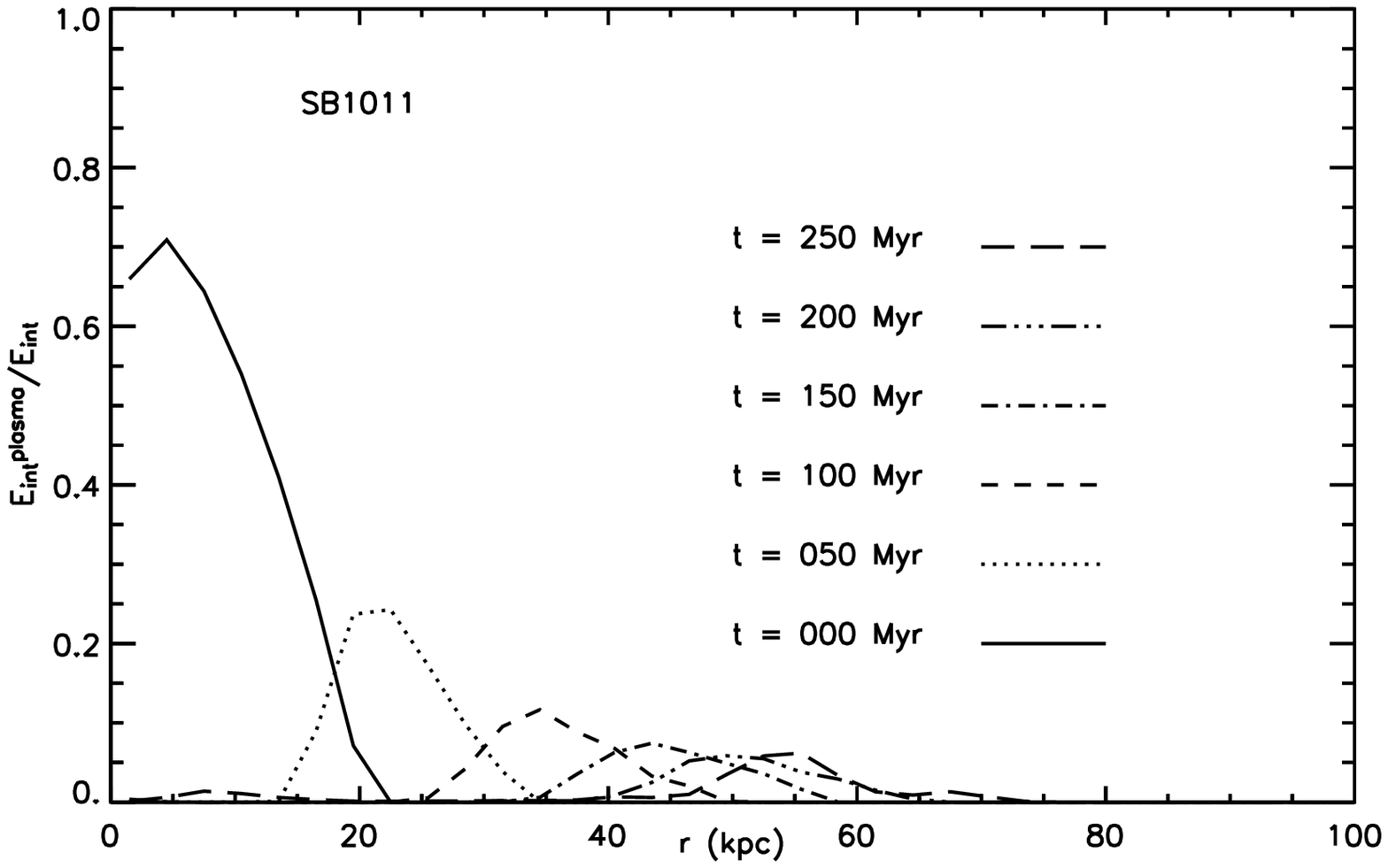}}
\caption{
Profiles of the fraction of the internal energy due to the bubble plasma.
The ratio is computed assuming that the fraction of plasma in each radial 
shell still hosts the same specific energy as in the initial conditions.}
\label{fig14}
\end{figure}

The use of 2D simulations enables us to extend this analysis up to the 
establishment of runaway cooling when some gas element reaches 
the mimimum temperature. In the case of the homogeneus cooling flow, we
saw that this happens soon after $t=390$ Myr. In SB1011, after a
turbulent oscillation, runaway cooling is established in the core
at some time after $t=500$ Myr. SB1020 presents an analogous behavior,
but because of the later infall, the cooling flow is more rapidly 
re-established and the runaway phase starts at $t=440$ Myr.
In the SB1030 case, the infalling material does not reach the center but 
rather compresses the material already present in the core; the effect is to
enhance the cooling so that the runaway phase begins after $t=380$ Myr.

\section{Conclusions}

We show the results of a set of 3D simulations of buoyant bubbles
in the atmosphere of a cool core cluster. We allow cooling, consider
the bubble plasma as relativistic, and achieve high effective resolution.
We consider the presented simulations to be a first step in studying
the interaction of plasma bubbles with the ICM. We will use these
runs as references for future work with different configurations 
of the system and additional physics. 

The dynamics of our bubbles resembles the results described 
by \citet{Saxton01} and by \citet{Reynolds05}. 
\citet{Saxton01} produce a set of 2D simulations similar to SB1030
for different values of the ratio of the density of plasma versus ICM.
They remark upon the robustness of the vortex rings and the fall of entrained
material to the core along the trunks; as in SB1030, this material
simply deposits onto the core rather than disrupts it.
\citet{Reynolds05} study the evolution of bubbles for different values 
of the ICM viscosity using 3D simulations. In a reference case without 
viscosity the bubble evolves similarly to SB1011, deforming 
in vortex rings with a thin trunk of material entrained in its wake.
Each of these studies focuses on the evolution of bubbles, rather than on their
impact on the environment, and therefore neglects cooling.

Our simulations do not fully address the problem of 
the quenching of the cooling flow by AGN, because the phase of
activity is neglected and many simplifications are assumed in the setup. 
We address, however, the direct effect of our bubbles on the cooling
and find that it is rather modest. Even in the SB1011 case, runaway cooling 
is postponed only by about 25\% respect to the unperturbed case. 
Setting the bubbles further away only reduces their effect, 
or even accelerates the cooling as in SB1030. 
Larger bubbles will probably have a bigger impact,
but the dimensions chosen in our cases are typical \citep{Birzan04}.

Numerically, the results are quite robust with respect to increase in resolution
and the adoption of different systems of coordinates. Thus we conclude
that our resolution is adequate. Details are different, however, and
2D simulations are much more sensitive to the propagation of rounding errors.
For this reason we prefer to trust the 3D cases and use 2D simulations only
to extend their results. To a lesser extent, 3D simulations suffer  
numerical instabilities as well, as revealed by the shredding of vortex rings 
along the lines of the computational grid.
One technique for mitigating grid effects involves just adding some random 
noise in the initial conditions, as in \citet{Jones05}. 
The symmetry of our results, however, testifies to the reliability of the code.

Future work will consider some relevant physics that is not included in
the present simulations. In particular, we will consider the magnetic field 
and the ICM viscosity, which have been shown to be able, for proper values,
to stabilize the bubble surface.
Using more realistic initial conditions, we will take into better account 
the activity phase of AGN and its effects on the medium.
The introduction of thermal conduction can be relevant in heating 
the entrained cold gas as it rises to regions of higher temperature,
besides directly heating the cluster core.
Mixing of hot and cold gas should also be properly treated,
as well as the transition of the plasma to the non-relativistic regime.
We aim to study also the interaction of pre-existing bubbles with shock waves 
induced by AGN activity or merger events \citep{Ensslin02,Heinz05}. 
Finally, a direct comparison with X-ray observations is possible using
tools like X-MAS \citep{Gardini04a}. 

\begin{acknowledgements}
The author acknowledges support under a Presidential Early Career Award
from the U.S. Department of Energy, Lawrence Livermore National Laboratory
(contract B532720) as well as the National Center for Supercomputing 
Applications. 
He wishes to thank Paul Ricker for his huge contribution to many aspects
of this work.
He also wishes to thank Robert Wilhelmson for useful reading suggestions.
FLASH was developed at the University of Chicago ASC Flash Center 
under support provided by Department of Energy grant number B341495.
\end{acknowledgements}

\end{document}